\include{graphicsx}

\documentclass{emulateapj}

\slugcomment{accepted to AJ on Jan. 26, 2007}

\shorttitle{Tight Ultracool Binary Discovered with Laser Guide Star Adaptive Optics}
\shortauthors{Siegler et al.}

\begin{document}

\title{Discovery of a 66 mas Ultracool Binary with Laser Guide Star Adaptive Optics} 
%\title{The Tightest Resolved Very Low-Mass Binary Discovered with a Ground-Based Telescope: 2MASS\,2132+1341AB$^{0}$} \footnotetext{Based on observations made with Keck Laser Guide Star Adaptive Optics.}

\author{Nick Siegler\altaffilmark{1}, Laird M. Close\altaffilmark{1}, Adam J. Burgasser\altaffilmark{2}, 
 Kelle L. Cruz\altaffilmark{3,4}, Christian Marois\altaffilmark{5}, Bruce Macintosh\altaffilmark{5}, \& Travis Barman\altaffilmark{6}}
\altaffiltext{1}{Steward Observatory, University of Arizona, Tucson, AZ 85721, USA}
\altaffiltext{2}{Massachusetts Institute of Technology, Kavli Institute for Astrophysics and Space Research, Cambridge, MA 02139, USA}
\altaffiltext{3}{Department of Astrophysics, American Museum of Natural History, New York, NY 10024, USA}
\altaffiltext{4}{NSF Astronomy and Astrophysics Postdoctoral Fellow}
\altaffiltext{5}{Institute of Geophysics and Planetary Physics L-413, Lawrence Livermore National Laboratory, Livermore, CA 94550, USA}
\altaffiltext{6}{Lowell Observatory, Flagstaff, AZ 86001, USA}

\email{nsiegler@as.arizona.edu}
\begin{abstract}
We present the discovery of 2MASS\,J21321145+1341584AB as a closely separated (0.066$\arcsec$) very low-mass field dwarf binary resolved in the near-infrared by the Keck II Telescope using laser guide star adaptive optics. Physical association is deduced from the angular proximity of the components and constraints on their common proper motion. We have obtained a near-infrared spectrum of the binary and find that it is best described by an L5\,$\pm$\,0.5 primary and an L7.5\,$\pm$\,0.5 secondary. Model-dependent masses predict that the two components straddle the hydrogen burning limit threshold with the primary likely stellar and the secondary likely substellar. The properties of this sytem - close projected separation (1.8$\pm$\,0.3\,AU) and near unity mass ratio - are consistent with previous results for very low-mass field binaries. The relatively short estimated orbital period of this system ($\sim$7-12\,yr) makes it a good target for dynamical mass measurements. Interestingly, the system's angular separation is the tightest yet for any very low-mass binary published from a ground-based telescope and is the tightest binary discovered with laser guide star adaptive optics to date.

\end{abstract}


%%%%%%%%%%%%%%%%%%%%%%%%%%%%%%%%%%%%%%%%%%%%%%%%%%%%%%%%%%%%%%%%%%%%%%%%%%%%%%%%%%%%%%%%%%%%%%%%%%%
\section{Introduction}

The coolest and lowest mass objects have historically been discovered as companions to low-luminosity stars. These objects include the two lowest luminosity spectral classes of low mass stars and brown dwarfs - the L and T dwarfs \citep[][and references within]{kir05}. The first L dwarf, GD\,165B, was discovered as the companion to a white dwarf \citep{bec88} while the first widely accepted brown dwarf, Gliese\,229B, was the companion to an M dwarf \citep{nak95,opp95}. Hence it is quite possible that the first of the ultracool brown dwarfs with effective temperatures less than $\sim$\,700\,K will also be discovered as a companion. These objects would likely populate a new spectral type beyond T with masses overlapping the planetary regime.

The hunt today for even cooler objects benefits from advances in high resolution imaging with the {\it Hubble Space Telescope} (HST) and large ground-based telescopes fitted with adaptive optics (AO).
With the spectral energy distributions of these cool objects peaking in the near-infrared (1-6\,\micron), observing at these wavelengths are advantageous for their detection and characterization. Thus observational strategies have relied on targetting continuously lower luminosity objects to further improve the contrast differential obtained in the near-infrared. 

An advantage in using AO over the HST is that they can be attached to larger ground-based telescopes attaining higher angular resolution and increased sensitivity to fainter sources. The challenge, however, exists in locating natural guide stars sufficiently bright ($R\lesssim$\,13.5\,mag, $K_s\lesssim$\,12\,mag) and near one's science target (isoplanatic angular distance $\lesssim$\,30$\arcsec$) to provide sufficient wavefront correction. There is less than 10\% chance of finding a natural guide star (NGS) meeting these requirements at 30\degr galactic latitude \citep{rod04}. The probability improves little even with the use of infrared wavefront sensors when targeting ultracool objects such as mid-L dwarfs (limiting magnitude of the NGS infrared wavefront sensor on NAOS at the Very Large Telescope is $K_s\sim$\,12\,mag\footnote{http://www.eso.org/instruments/naco/inst/naos.html}). Slightly better NGS sensitivity performance has been achieved using curvature wavefront sensors with avalanche photodiodes where $K_s\lesssim$\,12.3\,mag \citep{sie02}.

The search for substellar and planetary-mass objects through direct detection from ground-based telescopes now has a new technique - laser guide stars (LGSs). LGSs serve as artificial beacons for AO systems by exciting sodium atoms in the Earth's mesosphere at their resonant D-line frequency. These beacons serve as artificial (and steerable) guide stars which provide sufficient flux density for wavefront sensing and correcting. While LGS AO still requires a NGS to help correct both the lowest wavefront orders (``tip/tilt'', $\sim$\,2\,kHz) and the higher orders (``low-band wavefront sensor'', $\sim$\,0.01\,Hz), its flux density requirement is comparatively small ($R\,\lesssim$\,18\,mag). This results in $\sim$\,2/3 of the night sky accessible to high spatial resolution imaging \citep{liu06} and opens the door to probing the regions around ultracool L and T dwarfs never previously observed by ground-based telescopes. 

While the Keck II is the first of the 8-10\,m class telescopes to have an operational LGS AO system \citep{wiz06}, several more are expected to be commissioned within just the next 2 years \cite[see][]{liu06}. Several recent investigations using Keck II LGS AO have discovered companions to previously unresolved faint sources ushering in this new era of high-resolution imaging \cite[eg.][]{liu05,gel06,liu07,clo07}. 

In this investigation we observe six nearby ultracool\footnote{Generally defined as objects with spectral types later than M6 (T$_{eff}\lesssim$\,2700\,K); \cite{kir95,dah02}.} field dwarfs which we target for very faint companions. The objects were selected from the literature satisfying the following criteria: spectral type later than M6, never observed at high spatial resolution, too faint for current ground-based NGS AO systems, {\it and} spectrophotometric distances less than 30\,pc. We present here the discovery of one of the targets, 2MASS\,J21321145+1341584 \citep{cru07} as a closely-separated (0.066$\arcsec$) L dwarf binary resolved by the Keck II telescope NIRC2 infrared camera in combination with LGS AO. The binary is hereafter referred to as 2M\,2132+1341AB. The five other targets not found with near-equal mass companions are listed in Table \ref{tbl-1}. This discovery demonstrates the power of LGS AO - the ability to resolve a faint ($R$\,$\gtrsim$\,20\,mag; $J\sim$\,16\,mag) binary very near the diffraction limit (50\,mas) of a 10\,m telescope using an artificial beacon for wavefront correction.

%%%%%%%%%%%%%%%%%%%%%%%%%%%%%%%%%%%%%%%%%%%%%%%%%%%%%%%%%%%%%%%%%%%%%%%%%%%%%%%%%%%%%%%%%%%%%%%%%%%
\section{Observations and Data Reduction}
\subsection{Imaging}

The discovered binary system 2M\,2132+1341AB was observed on UT 2006 June 17 with the 10\,m Keck II telescope on Mauna Kea, Hawaii. It was the lone binary discovered from our sample of six ultracool dwarf targets. To optimize the resolution capabilities of our observations, we used the facility IR camera NIRC2 in the narrow (0.01\arcsec/pixel) camera mode with a 10$\arcsec\times$10$\arcsec$ field of view, in combination with the sodium LGS AO system \citep{buc04,wiz04}. All targets were observed through the broadband {\it K$_s$} filter (2.15\,$\micron$) where Strehl ratios are improved over {\it J} (1.25\,$\micron$) and {\it H} (1.63\,$\micron$). In the case of 2M\,2132+1341AB, observations were also made in both $J$ and $H$. All filters are of the Mauna Kea Observatories (MKO) filter consortium \citep{sim02,tok02}. Conditions were photometric for the majority of the night with better than 0.6$\arcsec$ seeing in the optical but with occasional windy periods.

Higher-order AO corrections were produced using the laser's on-axis light in the direction of the science target. This produced an emission similar to a {\it V}\,$\approx$\,10 point source. Lower order tip/tilt corrections were obtained using natural guide stars within 60$\arcsec$ of the targets. In the case of 2M\,2132+1341AB, the natural guide star used was 1036-0598908 ({\it R}\,=\,14.2\,mag) from the USNO-B1.0 catalog \citep{mon03}, located 13.6$\arcsec$ away. 

\begin{figure}
\includegraphics[angle=+0,width=\columnwidth]{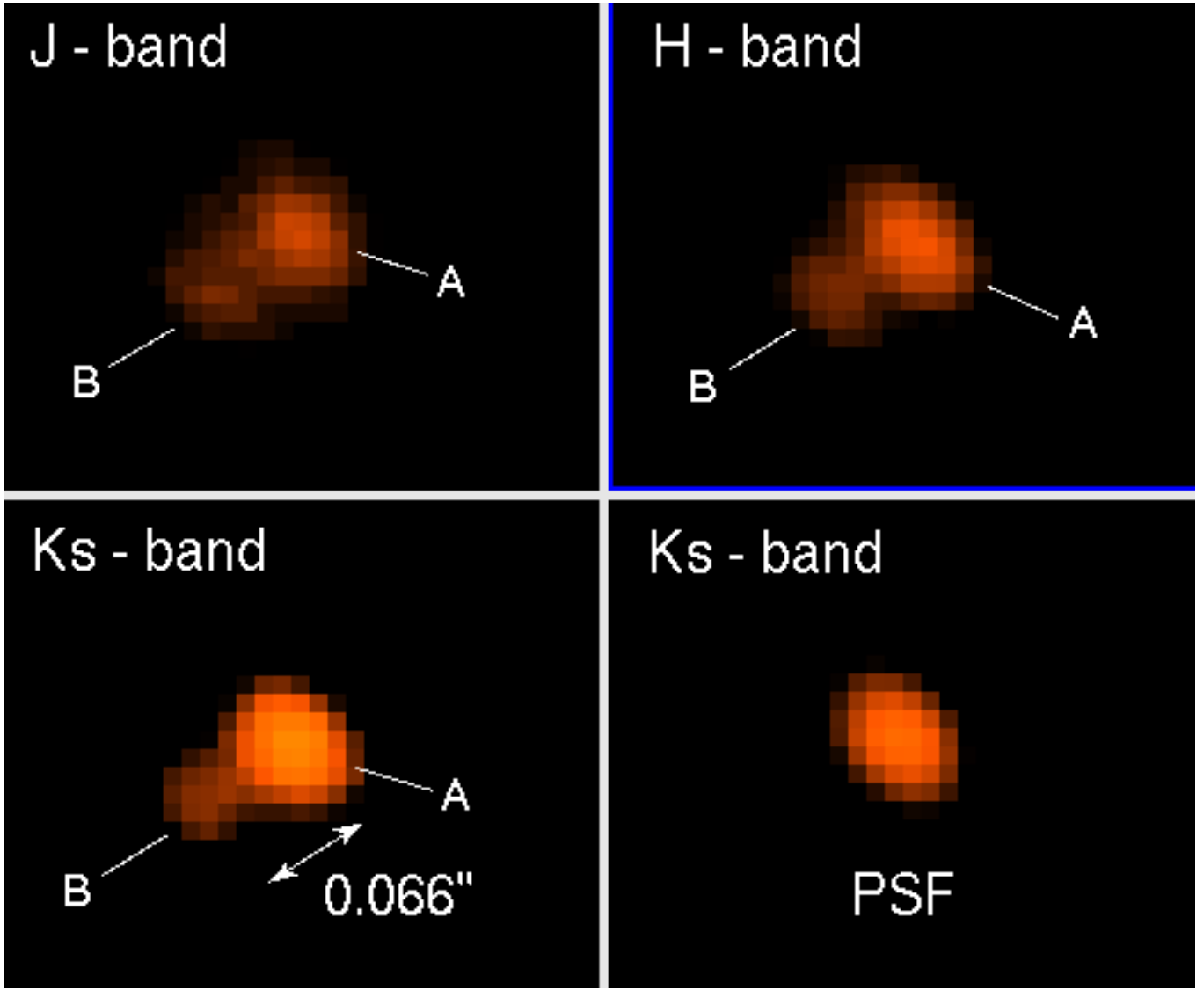}
\caption{{\it JHK$_s$}-band images of 2M\,2132+1341AB observed with Keck LGS AO; north is up and east to the left. We refer to the brighter component as the ``primary'' and designate it as 2M\,2132+1341A; the fainter component is referred to as the ``secondary'' or the ``companion'' and we designate it as 2M\,2132+1341B. The angular separation is only 66$\pm$4\,mas, among the tightest ultracool binaries ever resolved and the tightest yet resolved using a ground-based telescope. The LGS AO-corrected images have full width at half-maximum of 0.06$\arcsec$, 0.07$\arcsec$, 0.07$\arcsec$ at {\it J,H,} and $K_s$, respectively. Each image is 0.3$\arcsec$ on a side. The binary components are all slightly elongated along the telescope elevation axis (position angle $\sim$\,45\degr) believed to be due to telescope windshake. Also shown is one of the two PSFs used in the data reduction, 2MASS\,J16233609-2402209A \citep{clo07}, a young, stellar-mass M5 observed with similar elongation and air mass. 
\label{fig1}} 
\end{figure}

Table \ref{tbl-1} lists the five ultracool dwarf targets observed with no near-unity mass ratio companions detected at separations $\gtrsim$\,0.05$\arcsec$. Figure \ref{fig1} shows the resolved discovered binary 2M\,2132+1341AB. Both components are elongated along the telescope elevation axis projected to $\approx$\,45$\degr$, attributed to windshake during the observations \cite[a common problem with LGS AO on windy nights; see also][]{liu07}. The LGS AO-corrected images have full width at half-maximum of 0.06$\arcsec$, 0.07$\arcsec$, 0.07$\arcsec$ at {\it J,H,} $K_s$, respectively.

Each of the images shown in Figure \ref{fig1} was made by dithering a few arcsecs over three different quadrant positions on the NIRC2 narrow camera detector. Three images were taken per filter per dither position resulting in 2.5\,min for total on-source integration time per filter. The object was easily resolved into two components in all our data.

The images were reduced in a consistent manner using an AO data reduction pipeline written in the IRAF language as first described in \cite{clo02}. Modified for the NIRC2 narrow camera, the pipeline produces final unsaturated 15$\arcsec\times15\arcsec$ exposures in {\it J, H,} and $K_s$ with the highest signal-to-noise in the inner 5$\arcsec\times5\arcsec$ region. The photometric reduction pipeline uses the IRAF task ALLSTAR in the DAOPHOT point spread function (PSF) fitting photometry package\footnote{IRAF is distributed by the NOAO, which is operated by the AURA, Inc., under cooperative agreement with the NSF.}. The central regions of the pipeline's output is shown in Figure \ref{fig1}. 

Two different unsaturated single objects observed during the same night, but from a different program, were selected as PSF stars. These objects were observed with the same instrumental setup and showed similar Strehl ratios, FWHM, elongation due to windshake, and air mass. Both PSF sources, incidentally, are resolved primary objects of newly discovered wide (1.7$\arcsec$-1.9$\arcsec$) binaries. In both cases, the A and B components are sufficiently separated such that there is no flux contamination between them. The two PSFs used are 2MASS\,J16233609-2402209A, also shown in Figure \ref{fig1}, and 2MASS\,J16223020-2322240A \citep{clo07}. 2M\,2132+1341AB was fit independently with both PSF objects leaving behind clean residuals. The differential photometry in magnitudes is reported in Table \ref{tbl-2} and the photometric uncertainties are produced from the differences in the photometry between the two PSFs. These dominate the overall uncertainty.

\subsection{Spectroscopy}
Unresolved, low-resolution near-infrared spectroscopy of 2M\,2132+1341AB was obtained on 2005 October 17 (UT) using the SpeX
spectrograph mounted on the 3\,m NASA Infrared Telescope facility
\citep{ray03}. Conditions during the night were clear with
moderate seeing (0$\farcs7$--1$\farcs$0 at $J$-band). Data were
obtained utilizing the SpeX prism mode, providing a single-order
spectrum spanning 0.8-2.5\,$\micron$ spectrum with a dispersion of 20-30\,\AA/pixel. Use of the 0$\farcs$5 slit, aligned with the parallactic
angle, provided resolution $\lambda/\Delta\lambda$ $\approx$ 120 across
the near-infrared band. Six exposures of 2M\,2132+1341AB were obtained
in an ABBA dither pattern along the slit, with individual exposure
times of 150\,s.  The system was observed at an airmass of 1.02.  The
A0 star HD 210501 was observed immediately after the target exposures
at a similar air mass, followed by internal flat-field and Ar arc
lamps for pixel response and wavelength calibration.

     Data were reduced using the Spextool package, version 3.2
\citep{cus04}. The raw science data were processed by
performing linearity corrections, pairwise subtraction, and division
by a normalized flat field. The spectra were then extracted using the
Spextool default settings for point sources, and wavelength solutions
were calculated using the Ar arc calibration frames. Extracted spectra
from the same source were scaled to match the highest singal-to-noise spectrum of
the set, and the scaled spectra were median-combined. Telluric and
instrumental response features were removed following the procedure of
\citep{vac03}.

     The reduced spectrum of 2M\,2132+1341AB is shown in Figure \ref{fig2} ({\it black line}).
Strong absorption features of H$_2$O are present at 1.4 and 1.9\,$\mu$m, CO is prominent
at 2.3\,$\mu$m, and FeH is present at 0.99, 1.2, and 1.6\,$\mu$m. The
$J$-band spectral region exhibits a number of features that can be
attributed to K\,I and Na\,I lines in addition to FeH. There is no
indication of CH$_4$ in the spectrum of this source. These features
are all indicative of a late-type L dwarf, as also indicated by the
optical spectrum of \cite{cru07}.

%%%%%%%%%%%%%%%%%%%%%%%%%%%%%%%%%%%%%%%%%%%%%%%%%%%%%%%%%%%%%%%%%%%%%%%%%%%%%%%%%%%%%%%%%%%%%%%%%%%%%%%%%%%%%%%%%%%%%%%%%%
\section{Analysis}

 The key binary properties of 2M\,2132+1341A and B are derived here and summarized in Table \ref{tbl-3}. Individual apparent magnitudes are calculated from the observed $\Delta$ magnitudes (Table \ref{tbl-2}) and the integrated apparent magnitudes (unresolved) measured by the Two Micron All Sky Survey (2MASS) \citep{cut03}. Since the differential photometry observed with NIRC2 was measured with the MKO filter system, we converted the integrated {\it J} and {\it H} photometry from the 2MASS filter system to MKO using the color transformations of \cite{leg07}. While they provide no transformation between {\it K$_s$} MKO and {\it K$_s$} 2MASS, the transmission curves are very similar (1-2\% difference, S. Leggett private communication) and therefore we apply no correction. Uncertainties in the transformations and photometry are propagated in quadrature and reported in Table \ref{tbl-3}.

With measured differential photometry, derived apparent magnitudes, and a measured combined L6\,$\pm$\,0.5 optical spectrum from \cite{cru07}, what physical properties of the individual components can we infer? Since there is no known trigonometric parallax for the object, we rely on first estimating the component spectral types to derive absolute magnitudes using an empirical relation. This then enables estimates of the distance modulus, bolometric luminosities, and ultimately, with the aid of theoretical evolutionary tracks, masses and a period. The better constrained the component spectral types are, the more constrained (and meaningful) will the derived physical properties be. 
\subsection{Spectral Types}
The component near-infrared colors listed in Table \ref{tbl-2} by themselves provide only rough constraint on the individual spectral types \cite[eg.][]{chi06}. While the primary is certainly a mid-L dwarf, L3-L8, the possible spectral types for the secondary extend into the T-range, L3-T1. While the combined light spectrum is similar to that of
an L6, the components may have very discrepant spectral types.  The
secondary could even be a T dwarf without the characteristic CH$_4$
bands appearing in the combined light spectrum.  

To derive more precise estimates of the individual component spectral
types, we used a spectral synthesis technique based on that used by
\cite{bur09} to study L dwarf plus
T dwarf binaries\footnote{See also \cite{cru04,bur10,liu07,rei06}.}. A large sample
of composite spectra were generated by combining various pairings of
L5-T6 SpeX prism spectra obtained by A.\ Burgasser \& K.\ Cruz (72
individual spectra in all). The spectral types of the template spectra
are based on optical classifications for L dwarfs \citep[e.g.][]{kir99}
and near-infrared classifications for the T dwarfs \citep[e.g.][]{bur09}.
The components of these spectra were
constrained to have the same relative $K_s$-band magnitudes as measured
for 2M\,2132+1341AB, and to simultaneously be within 3$\sigma$ of the
measured $\Delta{J}$ (0.27\,mag) and $\Delta{H}$ (0.12\,mag).  The best
matches between the composite spectra and the observed (unresolved)
spectrum of 2M\,2132+1341AB were quantitatively determined by comparing
both relative $J$ and $H$ magnitudes and H$_2$O and CH$_4$ spectral
ratios \cite[defined in][]{bur11}. No assumption was made on the absolute
magnitudes of the individual components 
in this analysis so that the absolute magnitude/spectral type scale
was left as a free parameter.

\begin{figure}
\includegraphics[angle=+0,width=\columnwidth]{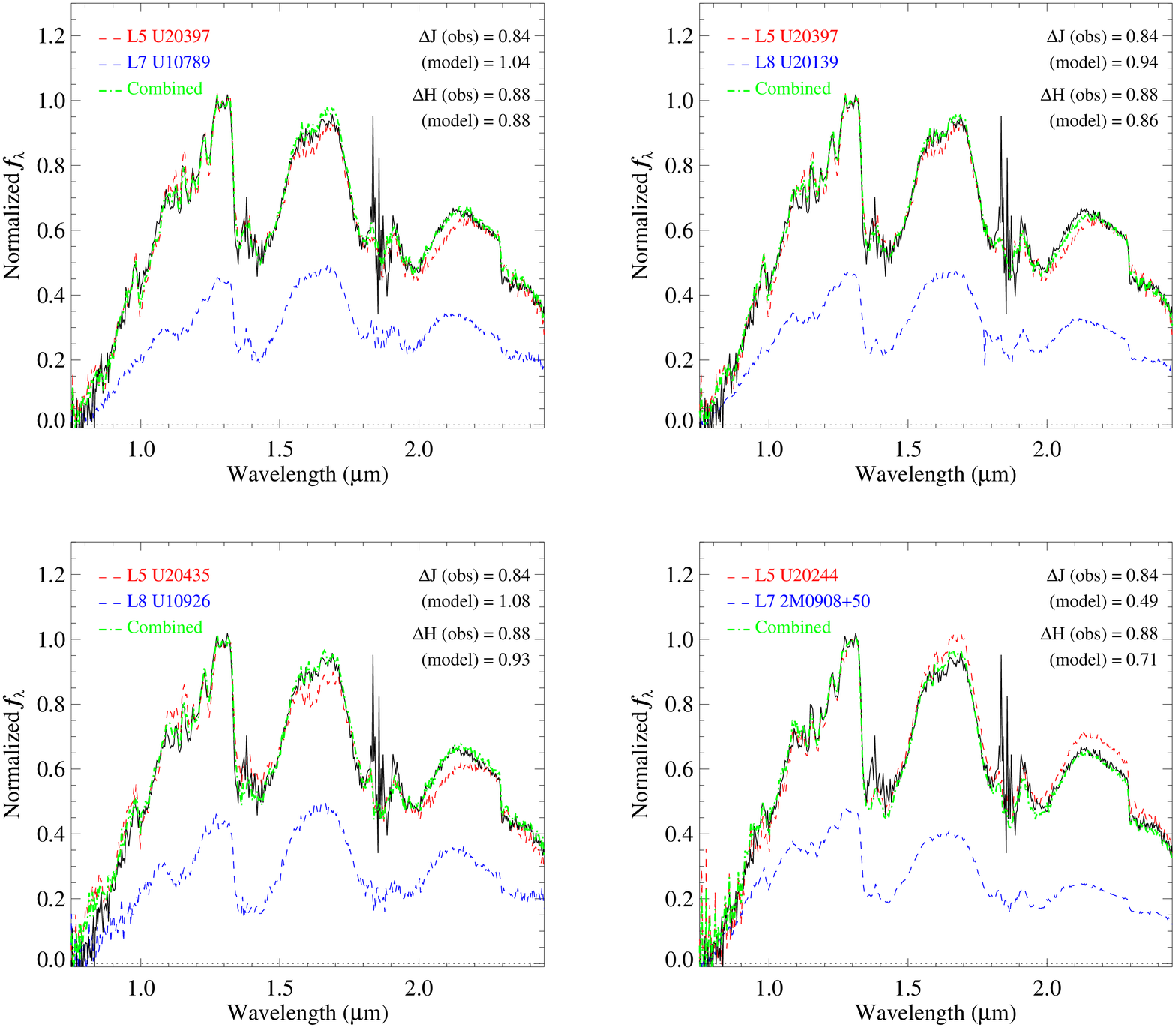}
\caption{Comparisons between the observed spectrum of 2M\,2132+1341AB and the composite spectra made by combining IRTF Spex spectra of spectral templates. In each panel, the observed near-infrared spectrum is represented by the {\it black line}. The upper ({\it red dashed line}) and lower spectra ({\it blue dashed}) in each panel are template guesses of the primary and secondary, respectively. The combined spectrum is represented by the {\it green dot-dashed line}. The {\it K} flux ratios between the secondary and primary are set to 2.29 ($\Delta\,K$=0.9\,mag) in each composite spectrum. The corresponding magnitude differences of the fits at $\Delta\,J$ and $\Delta\,H$ are shown in each panel. The two top panels and the bottom left suggest that an L5/L7.5 composite gives the closest match morphologically to the integrated 2M\,2132+1341AB spectrum at $J$ and $H$. We include the bottom-right panel to demonstrate that this kind of analysis is sufficiently robust to remove atypical component spectra (an atypically red L5 and an atypically blue L7) even when there is good morphological matching; see \S3.1.
% [{\it See the electronic edition of the Journal for a color version of this figure.}]
\label{fig2}} 
\end{figure}

Figure \ref{fig2} illustrates the three best-fit composite spectra based on
both the relative magnitudes and spectral ratio comparisons. In all three cases, an L5 spectral classification is selected for the primary along with an L7 or L8 for the secondary. In fact, this was the case for the best twenty fits. While the uncertainty is dominated by one subclass of uncertainties in the individual library spectral classifications, the consistency in the matches likely average out the overall uncertainty. Hence we conclude that the primary is a likely L5\,$\pm$\,0.5 and its companion a likely L7.5\,$\pm$\,0.5.

The fourth fit shown in Figure \ref{fig2} (bottom right) shows one of the combinations that was disqualified due to the disparity between the predicted and measured $\Delta$ magnitudes. The components of this system are
the unusually red L5 2MASS\,J062445.95-452154.8 (Reid et al., in prep.) and the
unusually blue L7 2MASS\,J09083803+5032088 \citep{cru03,cru07}. Despite a good morphological fit, this kind of analysis that includes $\Delta$ magnitudes as constraints to the properties of individual components is sufficiently robust to remove atypical component spectra. Of course, resolved near-infrared spectroscopy is required to verify the accuracy of these classifications.

\subsection{Physical Companions?}
Are the components of 2M\,2132+1341AB physical companions? Calculating spectrophotometric distances of the two sources separately results in equal values, 28\,$\pm$\,4\,pc. We used the fitted spectral types to independently obtain intrinsic flux densities (M$_K$) from the polynomial fit of \cite{bur07}. The distance's uncertainty includes those in the spectral
types and in a spectral type/absolute magnitude relation (see next section) taken in quadrature. In addition, assuming a surface density of order 10$^{-3}$ deg$^{-2}$ \citep{cru07} for all nearby L dwarfs, the probability of two lying within 0.1$\arcsec$ is $\approx$\,10$^{-7}$. Hence random alignment is very unlikely. Lastly, the 2M\,2132+1341 pair, or at least the primary, shows a large proper motion of 0.4$\arcsec$/yr \cite[NOMAD;][]{zac05}.
%\footnote{United States Naval Observatory Flagstaff Station (USNOFS) image and catalog archive database NOMAD \cite[Naval Observatory Merged Astrometric Dataset, http://www.nofs.navy.mil/data/fchpix;][]{zac05}. This database selects for each source the "best" astrometric and photometric data chosen from its catalogs and merges the results into a single dataset. For catalog details and references see http://www.nofs.navy.mil/nomad/nomad$\_$readme.html.}). 
A 2MASS {\it K$_s$} image of 2M\,2132+1341 observed in UT 1998 appears single. With sensitivity to point source brightness of K$_s$$\approx$15.3\,mag, 2M\,2132+1341B would have been detectable and resolved at separations $\gtrsim$\,1.5$\arcsec$ \citep{bur05}. Therefore we rule out 2M\,2132+1341B as an unrelated background object since its projected position nearly 8\,yrs ago would have been resolved in the 2MASS image. These factors provide strong evidence that the two sources are physical companions.

\subsection{Masses, Age, and Period}
With well-constrained spectral types in hand, we can now derive many of the binary's physical properties summarized in Table \ref{tbl-3}. An absolute $K$ magnitude for the primary is obtained using an M$_K$-spectral type relation from Figure 3 of \cite{bur07}, where binaries have been excluded. The companion absolute magnitude is then obtained by applying our measured $\Delta K_s$ ($\approx \Delta\,K$; S. Leggett, priv. comm.). Using our constrained component spectral types, we acquire the {\it K}-band bolometric corrections from \cite{gol04}, apply them to our $\Delta K$ photometry, and calculate the bolometric luminosity ratio between the components to be 0.32\,$\pm$\,0.08\,dex. Individual bolometric luminosities in units of solar luminosity, estimated from the component M$_K$ and BC$_K$ values, are 6.3$\times$10$^{-5}$\,$\pm$\,1.9$\times$10$^{-5}$\,L$_{\sun}$ for the primary and 3.0$\times$10$^{-5}$\,$\pm$\,1.0$\times$10$^{-5}$\,L$_{\sun}$ for the companion.

Individual masses of 2M\,2132+1341A and B can be estimated from theoretical evolutionary models using our derived bolometric luminosities and estimated ages of the system. The system's age, however, is less constrained. The binary does not appear affiliated with any moving-group or open cluster. Its optical spectrum shows no lithium or H$_{\alpha}$ spectral features \citep{cru07} suggesting that the source is more consistent with old field L dwarfs \citep{kir00,wes04}. Neither is there near-infrared color or optical spectrum evidence of sub-solar metallicity \cite[eg.][]{bur13,bur14} indicating the system is probably not a member of the
Galaxy's thick disk or halo populations \cite[$\gtrsim$\,10\,Gyr;][]{rei05}. In addition, the system's tangential motion of 53$\pm$\,2\,km\,s$^{-1}$ \cite[NOMAD;][]{zac05} is inconsistent with a young object ($\lesssim$\,1\,Gyr). 

\begin{figure}
\includegraphics[angle=+0,width=\columnwidth]{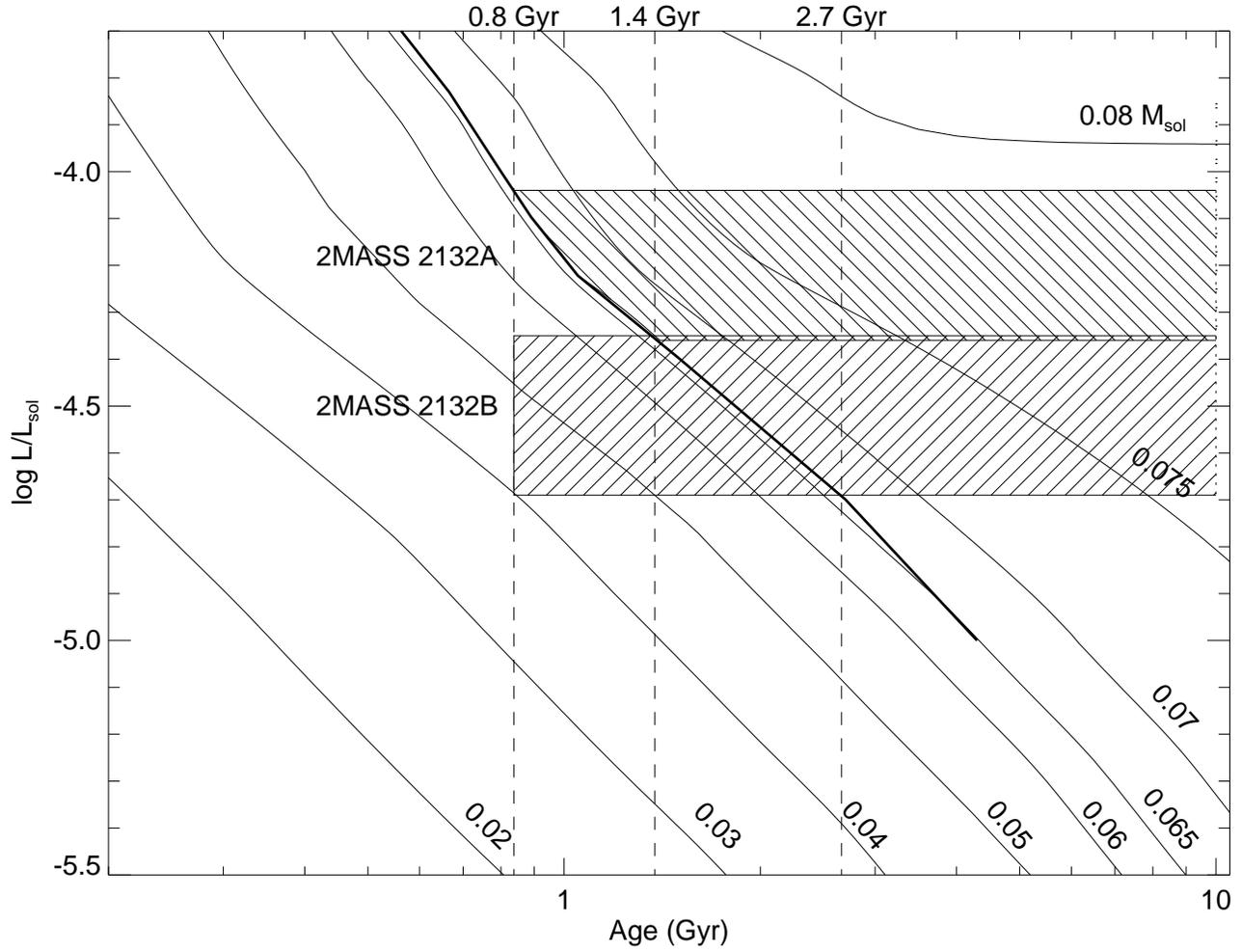}
%burgasser_L_age_figure.eps}
\caption{Theoretical evolutionary tracks from \cite{bur97} for 2M\,2132+1341A ({\it upper hatched region}) and B ({\it lower hatched region}). Diagonal {\it solid lines} show constant mass tracks; labelled numbers are in M$_{\sun}$ units. The stellar/substellar boundary for the model is $\approx$\,0.075\,M$_{\sun}$; the {\it bold solid line} represents the 1\% lithium depletion boundary drawn between the ages of 0.55-4.5\,Gyr. The lack of a lithium absorption feature in the combined optical spectrum \citep{cru07} suggests ages to the right of this line ($\gtrsim$\,0.8-1.3\,Gyr). Based on derived luminosity ranges and estimated upper-age limits discussed in \S3.3, the two {\it hatched} regions predict possible primary masses of 0.065-0.078\,M$_{\sun}$ and companion masses of 0.040-0.077\,M$_{\sun}$.
\label{fig3}} 
\end{figure}

The lack of lithium absorption in the optical spectrum can help place a lower mass limit to 2M\,2132+1341A of approximately 0.065\,M$_{\sun}$ depending on the system's age \citep{reb92,cha96,bas96,bur97}. Objects less than this limiting mass will always have central temperatures below the lithium-burning temperature. Slightly more massive objects will undergo lithium burning such that the element is observable at only younger ages. For example, using the models of \cite{bur97}, a 0.075\,M$_{\sun}$ object will undergo complete lithium burning in about 140\,Myr. In Figure \ref{fig3} we show their theoretical evolutionary tracks where we place the lower mass limit of 2M\,2132+1341A along a constant lithium abundance line of 1\% of the original abundance (similar to \cite{liu05}, we assume that a decrease in the initial lithium abundance by a factor of 100 marks the lithium absorption detection limit). This provides a lower age limit of 0.8-1.3\,Gyr, consistent with a weak or absent lithium absorption feature. Assuming the companion is coeval with the primary, this lower age along with the uncertainties in the secondary's luminosity predicts masses of 0.040-0.054\,M$_{\sun}$. A 10\,Gyr upper limit results in a primary mass of 0.077-0.079\,M$_{\sun}$ and a secondary of 0.076-0.077\,M$_{\sun}$. According to a theoretical analysis conducted by \cite{all05} of the age distribution of nearby field L dwarfs, there is a $\sim\,30\%$ probability that 2M\,2132+1341A and B are less than $\sim$\,1\,Gyr and a $\sim\,75\%$ chance that they are younger than $\sim$\,5\,Gyr. We list the median mass estimates for three ages in Table \ref{tbl-3} including a 5 Gyr best guess for stars in the solar neighborhood. Both objects likely straddle the hydrogen burning mass threshold \cite[$\approx$\,0.072-0.075\,M$_{\sun}$;][]{bur97,bar98} with the secondary most likely substellar.

The projected separation between the two components is only 1.8\,$\pm$\,0.3\,AU (at a distance of 28\,$\pm$\,4\,pc). We estimate the semimajor axis of 2M\,2132+1341AB by assuming that on average the true semimajor axis is 1.26 times larger than the projected separation \citep{fis92} or $<a>$\,=\,2.3\,AU. Using Kepler's third law and the range of possible masses, we estimate an orbital period of 7-12\,yr. Hence, this system is a good candidate target for astrometric
monitoring to derive orbital mass measurements \citep{lan01,bou04,zap04}.

%%%%%%%%%%%%%%%%%%%%%%%%%%%%%%%%%%%%%%%%%%%%%%%%%%%%%%%%%%%%%%%%%%%%%%%%%%%%%%%%%%%%%%%%%%%%%%%%%%%
\section{Discussion}

\subsection{How Typical are the Binary Properties of 2M\,2132+1341AB?}\label{bozomath}
VLM binaries are characterized by near-unity mass ratios (q$\sim$\,0.8-1.0) and tight separation distributions peaking between 3-10\,AU \cite[][and references within]{bur08}. According to the Very Low-Mass Binaries Archive\footnote{The Archive lists all the VLM binary systems reported in refereed journals, defined as binaries with total estimated mass less than $\sim$\,0.2\,M$_{\sun}$. This mass limit is arbitrary and corresponds to binary M6 field dwarfs (slightly earlier spectral types for younger objects). The website is maintained by Nick Siegler at http://paperclip.as.arizona.edu/$\sim$nsiegler/VLM$\_$binaries}, about a third of these systems are L/L binaries. The binary properties of 2M\,2132+1341AB, q\,$\gtrsim$\,0.9 and projected separation of 1.8\,$\pm$\,0.3\,AU, are consistent with these distributions.

Currently there are 16 known VLM binaries with angular separations less than the mean 66\,mas separation of 2M\,2132+1341AB. The tightest nine are spectroscopic binaries and are as yet unresolved. The subsequent seven were all discovered with the HST. Despite large aperture ground-based telescopes achieving AO corrected resolutions at {\it K} typically twice that of the HST, the space telescope's more stable point spread function allows for the identification of undersampled binaries. Interestingly, {\it 2M\,2132+1341AB is the tightest resolved very low-mass binary discovered by a ground-based telescope and the tightest using LGS AO}. The clear separation of this system
into two well-resolved components indicates that with good AO correction,
ground-based facilities can indeed achieve a superior resolution in the near-infrared
compared to HST.

\subsection{Future Dynamical Mass for 2M\,2132+1341AB}\label{bozomath}
Theoretical evolutionary models relating mass-luminosity-age relations are still largely uncalibrated for the lowest mass objects. In fact, only three VLM systems with constrained ages (all young) have had reliable orbits {\it and} resolved fluxes leading to derived individual kinematic masses - AB Dor C \cite[$\sim$\,50-100\,Myr;][]{clo05,luh05}, the eclipsing brown dwarf binary 2MASS\,J05352184-0546085AB found in Orion \cite[$\sim$\,1-2\,Myr;][]{sta06}, and GJ\,569Bab \cite[$\sim$\,500\,Myr;][]{zap04}. Unfortunately, neither 2M\,2132+1341AB's age nor distance is sufficiently well constrained to be used as a high accuracy luminosity-mass calibrator. However, if future high resolution optical spectroscopy (eg. repaired Space Telescope Imaging Spectrograph on HST) shows the presence of lithium in the companion, the system's age could be further constrained to $\sim$\,0.8-2.5\,Gyr making it a useful system for dynamical mass measurements. This would likely require, however, a widening in the components' projected separation. The HST and/or ground-base LGS AO observations should be able to measure significant orbital motion over the next $\sim$\,6\,yrs, similar to the study of 2MASSW\,J0746425+2000321 \citep{bou04}, the only dynamical mass measurement of an L dwarf binary.

%%%%%%%%%%%%%%%%%%%%%%%%%%%%%%%%%%%%%%%%%%%%%%%%%%%%%%%%%%%%%%%%%%%%%%%%%%%%%%%%%%%%%%%%%%%%%%%%%%%%%%%%%%%%%%%%%%%%%%%%%%
\section{Summary}
Keck II LGS AO observations of 2MASS\,J21321145+1341584 show that this very low-mass dwarf is a binary system. Observed differential near-infrared photometry and integrated spectra (optical and near-infrared) indicate that both components are consistent with mid-L dwarfs. Based on modeling the integrated optical spectra with spectra from 72 known L and T dwarfs, we identify 2M\,2132+1341A as an an L5\,$\pm$\,0.5 and 2M\,2132+1341B as an L7.5\,$\pm$\,0.5. The lack of lithium in the optical spectra suggests the primary's age is older than 800\,Myr. The system's very close separation (0.066$\arcsec$) and common proper motion from 2MASS infers a physical association. With conservative age estimate of 5\,Gyr, model-dependent masses suggest a system whose components straddle the hydrogen burning limit threshold with the primary likely stellar and the secondary likely substellar. The close projected separation (1.8$\pm$\,0.3\,AU) and near unity mass ratio of the system are consistent with previous results for field VLM binaries. The relatively short estimated orbital period of this system ($\sim$7-12\,yr) make it an ideal target for dynamical mass measurements. At the time of this writing, 2M\,2132+1341AB's angular separation is the tightest for any VLM binary discovered from a ground-based telescope and is the tightest using LGS AO.

%\end

%%%%%%%%%%%%%%%%%%%%%%%%%%%%%%%%%%%%%%%%%%%%%%%%%%%%%%%%%%%%%%%%%%%%%%%%%%%%%%%%%%%%%%%%%%%%%%%%%%%%%%%%%%%%%%%%%%%%%%%%%%%%%%%%%%%
\acknowledgements

The authors would like to acknowledge NASA and the NASA TAC for making this time available and the entire Keck LGS AO team for having set the bar for LGS performance. We also thank the referee Kevin Luhman for a thorough reading and suggested improvements. N.S. would like to thank Mike Cushing for discussions regarding L and T spectral classifications and Adam Burrows for providing model calculations. C.M. and B.M. note that their research was performed under the auspices of the U.S. Department of Energy by the University of California, Lawrence Livermore National Laboratory under contract W-7405-ENG-48, and also supported in part by the NSF Science and Technology Center for AO, managed by the University of California at Santa Cruz under cooperative agreement AST 98-76783. This research has made use of the Simbad and Vizier databases operated at CDS in Strasbourg, France; the Two Micron All Sky Survey (2MASS) data services, a joint project of the University of Massachusetts and the Infrared Processing Center/California Institute of Technology, funded by NASA and the NSF; the U.S. Naval Observatory (USNO) Naval Observatory Merged Astrometric Dataset (NOMAD). \textsc{iraf} is distributed by the National Optical Astronomy Observatories, which is operated by the Association of Universities for Research in Astronomy, Inc., under contract to the NSF.

%%%%%%%%%%%%%%%%%%%%%%%%%%%%%%%%%%%%%%%%%%%%%%%%%%%%%%%%%%%%%%%%%%%%%%%%%%%%%%%%%%%%%%%%%%%%%%%%%%%%%%%%%%%%%%%%%%%%%%%%%%%%%%%%%%%

%%%%%%%%%%%%%%%%%%%%%%%%%%%%%%%%%%%%%%%%%%%%%%%%%%%%%%%%%%%%%%%%%%%%%%%%%%%%%%%%%%%%%%%%%%%%%%%%%%%%%%%%%%%%%%%%%%%%%%%%%%%%%%%%%%%%
\clearpage
\begin{deluxetable}{lccc}
%\tabletypesize{\scriptsize}
\tabletypesize{\tiny}
\tablecaption{Ultracool Field Dwarfs Observed with No Physical Companion Detections\tablenotemark{a}  \label{tbl-1}}
\tablewidth{0pt}
\tablehead{
\colhead{2MASS Name} &
\colhead{{\it K$_s$}} &
\colhead{Spectral Type} &
\colhead{References} \\
}

\startdata
LSR\,J1610-0040  & 12.02 & sdM:sdL: & 1,2\\
2MASS\,J17210390+3344160	          & 12.47 & L3 & 3\\
SDSS\,J202820.32+005226.5                   & 12.79 & L3 & 4\\
2MASS\,J20343769+0827009\tablenotemark{b} & 13.08 & M:L & 5\\
2MASS\,J22490917+3205489\tablenotemark{c}  & 13.59 & L5 & 6\\
\enddata
\tablenotetext{a}{For near-equal mass binaries (mass ratio $\gtrsim$ 0.7), the angular separation sensitivity is $\sim$50\,mas. For less massive companions (q\,$\lesssim$\,0.7), sensitivity improves with increasing angular separation up to our observation's 10\arcsec radial field of view.}
\tablenotetext{b}{Object was originally classified as a mid-L dwarf but due to insufficient signal-to-noise ratio is now only roughly estimated as a late-M/early-L dwarf.}
\tablenotetext{c}{A faint point source at PA=194$\degr$, separation $\sim$\,90\,mas was observed in $J, H$, and $K_s$ but determined to be a ``super-speckle'' due to its wavelength-dependent angular separation.}
%\tablecomments{Epoch 2000}
\tablerefs{(1) \cite{lep03}, (2) \cite{cus06}, (3) \cite{cru03}, (4) \cite{haw02}, (5) Kelle Cruz, priv. comm., (6) \cite{cru07}}
\tablecomments{Each target has at least one bright NGS $\lesssim$\,30$\arcsec$ serving as the tip/tilt and low-band source.}
\end{deluxetable}

%\clearpage
\begin{deluxetable}{lc}
%\tabletypesize{\scriptsize}
\tabletypesize{\tiny}
\tablecaption{2M\,2132+1341AB Observed Properties \label{tbl-2} }
\tablewidth{0pt}
\tablehead{
\colhead{Property} &
\colhead{Measurement} \\ 
}
\startdata
 $\Delta\,J$  &  0.84\,$\pm$\,0.09\,mag \\
 $\Delta\,H$  &  0.88\,$\pm$\,0.04\,mag \\
 $\Delta\,K_s$  &  0.90\,$\pm$\,0.04\,mag \\
$J_A$ & 16.07\,$\pm$\,0.07\,mag\\
$J_B$ & 16.91\,$\pm$\,0.12\,mag\\
$(J-K_S)_A$ & 1.84\,$\pm$\,0.09\,mag\\  
$(J-K_S)_B$ & 1.78\,$\pm$\,0.14\,mag\\ 
$(J-H)_A$ & 1.04\,$\pm$\,0.09\,mag\\  
$(J-H)_B$ & 1.00\,$\pm$\,0.13\,mag\\  
Separation & 66\,$\pm$\,4\,mas \\
Position angle & 121.94\,$\pm$\,1.30$\degr$ \\
Date observed & UT 2006 Jun 17 
\enddata
\tablecomments{Photometry on the MKO system.}
\end{deluxetable}

%\clearpage
\begin{deluxetable}{lc}
\tabletypesize{\scriptsize}
\tablecaption{2M\,2132+1341AB Derived Properties \label{tbl-3}}
\tablewidth{0pt}
\tablehead{
\colhead{Property} &
\colhead{Value}}
\startdata
Spectral Types & \\
\,\,\,\,\,\,A + B (optical) & L6\,$\pm$\,1 \\
\,\,\,\,\,\,A & L5\,$\pm$\,0.5 \\
\,\,\,\,\,\,B & L7.5\,$\pm$\,0.5 \\
$M_{K_A}$  &  11.92\,$\pm$\,0.33\,mag \\
$M_{K_B}$  &  12.82\,$\pm$\,0.30\,mag  \\ 
Distance & 28\,$\pm$\,4\,pc\\
Luminosities:\\
\,\,\,\,\,\,L$_A$ & 6.3$\times$10$^{-5}$\,$\pm$\,1.9$\times$10$^{-5}$\,L$_{\sun}$\\
\,\,\,\,\,\,L$_B$ & 3.0$\times$10$^{-5}$\,$\pm$\,1.0$\times$10$^{-5}$\,L$_{\sun}$\\
%\,\,\,\,\,\,log\,[L$_A$/L$_{\sun}]$ & -4.20\,$\pm$\,0.14\,dex\\
%\,\,\,\,\,\,log\,[L$_B$/L$_{\sun}]$ & -4.52\,$\pm$\,0.14\,dex\\
%Mass ratio (log M$_B$/M$_A$) & 0.78\,$\pm$\,0.13\\
Masses (A/B): & \\
%; model uncertainty $\sim$\,\pm\,5\,M$_{Jup}$:
\,\,\,\,\,\,0.8\,\,Gyr & 0.065/0.048\,M$_{\sun}$\\
\,\,\,\,\,\,5\,\,Gyr & 0.077/0.075\,M$_{\sun}$\\
\,\,\,\,\,\,10\,\,Gyr & 0.078/0.076\,M$_{\sun}$\\
Proper motion (NOMAD): & \\
\,\,\,\,\,\,  $\mu_{\alpha}$cos$\delta$ & -55.3\,$\pm$\,9.0\,mas\,yr$^{-1}$\\
\,\,\,\,\,\,  $\mu_{\delta}$ & -394.7\,$\pm$\,9.0\,mas\,yr$^{-1}$\\
Separation (projected)     & 1.8$\pm$\,0.3\,AU \\
Orbital period & 7-12\,yr\\
%\hline\hline
\enddata
%\tablenotetext{a}{MKO system}
\tablecomments{(1) All photometry on the MKO filter system, (2) See \S3 for details and references.}
\end{deluxetable}

\clearpage

\clearpage

\clearpage

\end{document}